%% file: main.tex
\documentclass[11pt]{article}
\pdfoutput=1
\usepackage[utf8]{inputenc}
\usepackage[left=1in,right=1in,top=1in,bottom=0.7in]{geometry}
\usepackage{microtype}
\usepackage{amsmath,amssymb}
\usepackage{graphicx}
\usepackage{booktabs,tabularx,array}
\usepackage{caption,subcaption}
\usepackage{setspace}
\usepackage{placeins}
\usepackage{xcolor}
\usepackage[numbers,square]{natbib}
\usepackage{xurl}
\definecolor{linkblue}{RGB}{25,76,153}
\usepackage[colorlinks=true,citecolor=linkblue,linkcolor=linkblue,urlcolor=linkblue]{hyperref}
\usepackage{subcaption}
\usepackage{svg}
\usepackage{cleveref}

\renewenvironment{abstract}{%
  \par\begin{center}\bfseries Abstract\end{center}%
  \begingroup\setstretch{1.05}\setlength{\parindent}{0pt}\noindent\ignorespaces
}{\par\endgroup}

\title{\textbf{Perceived Agency During Agent–Avatar Control Transitions in Virtual Reality}}
\author{
Jingyi Zhang\textsuperscript{1},
Ruijun Sun\textsuperscript{1},
Klara Brandstätter\textsuperscript{1},
Nels Numan\textsuperscript{1},
Anthony Steed\textsuperscript{1},\\[0.5em]
\normalsize \textsuperscript{1}University College London
}

\date{\today}

\begin{document}
\newgeometry{left=1in,right=1in,top=0.6in,bottom=0.7in}
\begingroup
\setstretch{1}
\maketitle
\endgroup
\vspace{-0.5em}
\begin{abstract}
Prior beliefs about whether a virtual human (VH) is controlled by a human or a computer can shape how users interpret and respond to its behaviour. As emerging hybrid VH systems allow control to transfer between human operators and computer programs during an ongoing interaction, it remains unclear whether perceived agency updates with the actual controller or continues to reflect users’ prior beliefs and first impressions. We investigated this question in a within-subject VR study with 24 participants across four conditions: always computer-controlled, computer-to-human, human-to-computer, and always human-controlled. We designed an interaction task in which participants acted as interviewers, asking the VHs a fixed set of questions. Before each interaction, participants were informed whether the VH was human- or computer-controlled, matching the actual controller at the beginning, but were unaware of possible control transitions.

We found that perceived agency was strongly influenced by the controller presented at the beginning of the interaction, with no evidence of updating following the control transitions. Social evaluations including likeability and co-presence were generally higher in conditions where human control occurred later and covered a greater proportion of the interaction. The results suggest that prior beliefs and first impressions continued to shape perceived agency, whereas social evaluations were more sensitive to the VH’s actual behaviour over the interaction.
\end{abstract}

\vfill

\clearpage

\clearpage
\restoregeometry

\input{1.Introduction}

\input{2.Related_Work}

\input{3.Method}

\input{4.Result}

\input{5.Discussion}

\input{6.Conclusion}

\clearpage
\begingroup
\singlespacing
\small
\phantomsection
\addcontentsline{toc}{section}{References}
\bibliographystyle{abbrv-doi}
\bibliography{template}
\endgroup

\clearpage
\section*{Supplementary Information}
\addcontentsline{toc}{section}{Supplementary Information}
\setcounter{section}{0}
\setcounter{subsection}{0}
\setcounter{table}{0}
\setcounter{figure}{0}
\renewcommand{\thesection}{S\arabic{section}}
\renewcommand{\thetable}{S\arabic{table}}
\renewcommand{\thefigure}{S\arabic{figure}}
\input{supplement}

\end{document}

%% file: 1.Introduction.tex
\section{Introduction} 


Prior beliefs about interaction partners can shape how people interpret those partners' behaviour and respond to them. The same message, for example, may be perceived as more helpful and comforting when it comes from a close friend than from a casual acquaintance \cite{clark_impact_1998}. 
In computer-mediated communication, where interaction takes place through text, voice, avatars, or other technological interfaces, social cues may be filtered through the medium or presented by the technology itself \cite{xu_explicating_2020}. Although people can still form impressions from the cues that the medium makes available \cite{walther_impression_1993}, when these cues are ambiguous, people may rely more on prior expectations and stereotypes to fill in missing information and interpret the message \cite{epley_when_2005}. In such settings, what people believe about the source of the communication becomes important.

With advances in artificial intelligence (AI), computer systems can now generate fluent text messages \cite{jones_people_2025}, engage in audio conversations \cite{kyrlitsias_social_2022}, and generate human movements and nonverbal behaviours for virtual humans (VHs) \cite{zhu_human_2024, kyrlitsias_social_2022}. In some circumstances, particularly text-based communication, it is already becoming difficult to determine whether one is communicating with a human or an AI agent \cite{jakesch_human_2023, jones_people_2025}. As these technologies continue to advance, 
it will become increasingly difficult to tell whether a VH is controlled by a human or an AI.

However, attitudes and behaviours towards human and AI partners can differ even when the content they produce is identical. Recipients felt less heard when a response was attributed to an AI agent rather than to a human \cite{yin_ai_2024}. When the same virtual interviewer was believed to be computer-controlled, participants were more willing to disclose information than when it was believed to be human-controlled \cite{lucas_its_2014}. Therefore, it is important to understand how beliefs about whether an interaction partner is human- or computer-controlled shape social interactions.

Immersive virtual environments provide a distinctive setting to investigate this process. Unlike text-based communication, interaction in virtual reality (VR) can convey rich social cues, both verbal and non-verbal. At the same time, the appearance of a VH does not necessarily reveal who or what controls it. 
Previous work has shown that people respond socially to VHs believed to be either human- or computer-controlled \cite{nowak_effect_2003,felnhofer_virtual_2024}. Therefore, VR is particularly suitable for studying how beliefs about a VH's controller shape social interaction.

Following established terminology, we refer to a human-controlled VH as an \emph{avatar} and a computer-controlled VH as an \emph{agent} \cite{felnhofer_virtual_2024}. \emph{Perceived agency} is defined as ``the extent to which individuals perceive virtual others as representations of real persons in real time'' \cite{blascovich_theoretical_2002}. We use \emph{actual agency} to refer to the actual controller behind a VH and \emph{disclosed agency} to refer to the controller presented to participants as controlling the VH. Actual and perceived agency may differ \cite{felnhofer_virtual_2024}, and disclosed agency may differ from either or both.

Previous research has reported mixed findings on the effects of disclosed and perceived agency. Von der Pütten et al. \cite{von_der_putten_it_2010} found no evidence that disclosed agency affected participants' evaluations of the VH or their behavioural responses, whereas behavioural realism affected both. A recent meta-analysis found that VHs perceived as avatars elicited higher social presence and more favourable evaluations than VHs perceived as agents, while no effects were observed for behavioural outcomes \cite{felnhofer_virtual_2024}.

Most studies manipulate perceived agency through agency disclosures at the beginning of a session \cite{felnhofer_virtual_2024}, 
and treat perceived agency as a static variable. However, as an interaction unfolds and further behavioural cues accumulate, does perceived agency remain stable? Research on social belief updating shows that unexpected behaviour can lead people to revise beliefs about a partner's intentions \cite{christian_belief_2024}. 
However, if a prior belief is strong, people may instead explain inconsistent behaviour without substantially changing that belief \cite{kim_psychology_2020}.
Which of these applies to perceived agency? Will it be updated when a VH's behaviour becomes inconsistent with its disclosed agency, or will the initial agency attribution persist?

These questions become more important as hybrid VH systems that allow control to pass between computer algorithms and human operators arise. For example, Zhang et al. \cite{zhang_single_2025} developed a dynamic VH control system in which multiple VHs perform pre-recorded behaviours during non-interactive phases, while a human operator can take control of any of them to deliver more complex interactions. The authors claimed that the system has potential applications in VR shows and complex training scenarios for teachers or doctors. Beyond VHs that represent fictional characters, advances in technology now allow the creation of AI clones, digital replicas of real humans that can mimic their voices, appearances, and behavioural patterns \cite{lee_speculating_2023}. Building on this, Leong et al. \cite{leong_dittos_2024} proposed Dittos, personalized embodied agents designed to represent people in virtual meetings by visually resembling them, replicating their voices, and incorporating personal knowledge. When integrated with hybrid control systems, such personalized agents could create new interaction paradigms but at the same time introduce ethical concerns. However, in those hybrid systems, can a seamless transition between controllers leave perceived agency unchanged, or will users update their perceived agency when control changes?




Therefore, we investigate how perceived agency develops and how social responses are affected when the actual controller of a VH changes during an ongoing interaction. We conducted a within-subject experiment with four conditions: two consistent agency conditions (always agent or always avatar) as baselines, and two dynamic agency conditions involving a control transition (agent-to-avatar or avatar-to-agent). To keep participants focused on the VH, we designed an interview task that required participants to act as an interviewer, asking a VH pre-defined questions. This study was conducted in two distinct phases. In the first phase, participants were unaware of any potential control transition, which allowed us to observe their natural behaviour and perceptions during live interaction. They were told at the beginning of each condition that they were going to interact with a human-controlled VH or a computer-controlled VH. The disclosed agency was aligned with the initial controller regardless of whether a control transition occurred. We collected objective measures including gaze and head movement, and subjective measures of perceived agency, social presence, and likeability. In the second phase, they were informed of the possibility of a control transition and were asked to review a video of their previous interactions, using a slider to indicate their perceived agency. They were asked to think aloud while adjusting the slider, and more detailed qualitative responses were collected through interviews after they reviewed each session.


We found no evidence that perceived agency changed following either direction of control transition. Participants' judgements remained strongly structured by their initial agency attribution but leant towards human control in both transition conditions, although they interpreted local behavioural cues as evidence for human or computer control. At the same time, interactions involving longer or later human control were evaluated as more likeable, consistent with the co-presence results, suggesting that evaluations of the interaction may respond to controller behaviour without producing an explicit reclassification of agency.

Our contributions include a two-phase, mixed-methods study providing empirical evidence that perceived agency can remain anchored to a prior belief despite a change in actual control during an interaction, whereas social evaluations appear more sensitive to the VH’s behaviour over the interaction. We also provide insights from participant interviews into how behavioural cues shape perceptions of agency in hybrid agent–avatar interactions.





%% file: 2.Related_Work.tex
\section{Related Work}
\subsection{Agency}
In VR, the perceived agency, which refers to the agency as experienced by the user, can differ from the actual agency that reflects the controller behind the VH \cite{felnhofer_virtual_2024}. When interacting through avatars in VR, users may not always be able to distinguish whether they are engaging with a human or a computer algorithm \cite{nowak_avatars_2018}. 

Previous studies have manipulated perceived agency by shaping participants’ impressions through verbal instructions or by introducing confederates who were presented as co-players \cite{kothgassner_real-life_2017,felnhofer_mere_2019, von_der_putten_it_2010}. In many studies, perceived agency was manipulated by changing what participants were told while the VH itself remained computer-controlled \cite{felnhofer_virtual_2024, fox_avatars_2015}. Moreover, prior studies generally manipulate agency across different sessions rather than changing it within a single ongoing interaction. As a result, it remains unclear whether perceived agency can change dynamically when the actual controller switches during the same interaction session.


Although both avatars and agents can elicit social responses \cite{nowak_effect_2003}, prior research has shown that perceived high agency (avatar) produced stronger social influence than perceived low agency (agent) \cite{fox_avatars_2015}. Among desktop studies, stronger agency effects were found for objective than for subjective measures. Agency effects were also stronger in competitive and cooperative tasks than in neutral tasks, and when VHs in both conditions were controlled by actual humans rather than by computer algorithms. However, von der Pütten et al. \cite{von_der_putten_it_2010} reported that participants’ evaluations of the VH and their behavioural responses differed little across disclosed-agency conditions, whereas behavioural realism affected both. In VR, a recent meta-analysis found higher social presence and more favourable evaluations for perceived avatars than for perceived agents, but no reliable differences in behavioural outcomes \cite{felnhofer_virtual_2024}.



\subsection{Behavioural Realism}
Behavioural realism describes the extent to which a VH behaves as an actual person would \cite{kyrlitsias_social_2022}. Prior research has shown that behavioural realism can influence users' sense of social presence \cite{kyrlitsias_social_2022}, perceptions of agency \cite{pfeiffer_non-verbal_2011}, and evaluations of the VH and behavioural responses \cite{von_der_putten_it_2010}. Increasing behavioural realism means making the VH behave more like a human during face-to-face interaction. Behavioural realism could be determined by many factors, including verbal and non-verbal behaviours, such as body movements, gestures, facial expressions, and gaze, as well as responsiveness and interactivity with the environment and the user \cite{kyrlitsias_social_2022}. We refer back to behavioural realism in 
\cref{sec: discuss behavioural realism} to consider how differences between conditions may have shaped perceived agency and social evaluations.

\subsection{Social Response Models} \label{sec: related work social response models}
According to Felnhofer et al. \cite{felnhofer_virtual_2024}, the two leading theories of social responses and perceived agency are the Threshold Model of Social Influence \cite{biocca_toward_2003, blascovich_theoretical_2002} and the Media Equation Concept \cite{nass_machines_2000}. The former implies that perceived agency and behavioural realism jointly shape how users respond to virtual entities. That is, higher behavioural realism is required for a virtual human to elicit social responses when users perceive they are interacting with an agent. Conversely, lower behavioural realism is needed when users believe they are interacting with an avatar. The latter suggests that social reactions arise regardless of perceived agency, which means users respond the same way to both agents and avatars. Empirical studies support both theories, with evidence supporting the former \cite{guadagno_virtual_2007, doi:10.1177/0146167203029007002} and the latter \cite{KothgassnerDoes, von_der_putten_it_2010}. In \cref{sec: model discuss}, we refer back to these models to consider which one our results follow.

\subsection{Measures of Social Responses}
Prior research on agency has examined social responses using both objective and subjective measures. Objective indicators typically include low-level behaviours, such as gaze patterns, and physiological signals like heart rate and electrodermal activity \cite{felnhofer_virtual_2024}. Subjective responses are usually measured through self-report questionnaires assessing factors such as likeability of the virtual interaction partner \cite{hoppe_human_2020} and social presence \cite{kothgassner_agency_2018, von_der_putten_it_2010, harms_internal_2004}. Social presence, defined as the feeling of ``being with another'' \cite{biocca_toward_2003}, is important, as greater social presence during interactions with VHs is associated with social reactions closer to those in face-to-face interactions \cite{kyrlitsias_social_2022}. 

%% file: 3.Method.tex
\section{Methods}
To investigate participants' perceptions of agency while interacting with VHs, we conducted a two-phase, within-subject experiment. The first phase (\textit{Phase I}) collected behavioural and perceptual data during participants' interactions with the VHs, with participants unaware of any potential change in the controller behind the VH. The second phase (\textit{Phase II}) gathered retrospective data as participants watched recordings of their initial interactions after being informed about the possibility of a control switch. The study was run under ethics approval by the UCL Computer Science Research Ethics Committee (approval ID UCL/CSREC/R/16).

\subsection{Participants}
We recruited 24 participants (10 male) through our lab's mailing list, lab's open day, and posters placed in the building where our lab is located. Participants' ages ranged from 18–24 to 35–49 years (mode: 25–29 years, n = 12). All had normal or corrected-to-normal vision. Each participant received a £15 Amazon voucher as compensation. Each participant attended a single 75-minute session and completed \textit{Phase I} followed immediately by \textit{Phase II}.

\subsection{Tasks} \label{sec:tasks}
In \textit{Phase I}, to keep participants focused on the VH's behaviour and responses, we designed a structured interview task in which participants acted as interviewers and the VH they interacted with served as the interviewee. To ensure consistency and control across participants, they were allowed to ask only the following pre-defined questions in a fixed order, with no follow-up questions. \textbf{Q1:} Can you introduce yourself? \textbf{Q2:} What's something you're good at? \textbf{Q3:} Can you think of a little problem you solved recently? \textbf{Q4:} How do you usually plan your day or tasks? \textbf{Q5:} What's something new you've learned lately? 

In \textit{Phase II}, participants were informed that the controller might transition between a human and a computer during an interaction, but that a transition would not necessarily occur. While watching recordings of their interactions during \textit{Phase I}, they continuously adjusted a slider with one end labelled ``Real Human'' and the other labelled ``Computer Program'' to indicate who they believed was controlling the VH at each point in the video. Participants were also asked to think aloud while completing this task.

\subsection{Experimental Design}
The experiment employed a within-subject design with four conditions: 

\begin{itemize}
    \item \textbf{Condition 1 - Always Agent:} Participants were led to believe they were interacting with an agent, and control remained with the agent throughout the interaction. 
    \item \textbf{Condition 2 - Agent to Avatar:} Participants were led to believe they were interacting with an agent. During the interaction, control switched to a human.
    \item \textbf{Condition 3 - Avatar to Agent:} Participants were led to believe they were interacting with a human. Control then switched to an agent during the interaction.
    \item \textbf{Condition 4 - Always Avatar:} Participants were led to believe they were interacting with a human, and control remained with the human throughout the interaction.
\end{itemize}

We counterbalanced the order of conditions, virtual human appearances, and their responses using mutually orthogonal Latin squares (MOLS) \cite{10.5555/1202540}. Specifically, we implemented three 4$\times$4 MOLS, repeated across six four-participant blocks (N=24), ensuring that each condition, virtual human appearance, and response set appeared once per position and that all Condition $\times$ Appearance, Condition $\times$ Response, and Appearance $\times$ Response pairings were equally represented. The full order is in Supplementary Materials.


Before each interaction, participants were informed by the second experimenter whether they were going to interact with a human- or computer-controlled VH. Prior to all the interactions, participants were introduced via a Microsoft Teams video call to a confederate (the lead experimenter) whom they were told might control the VH.

The control transition always occurred after the VH finished responding to the second question and before the VH began answering the third question. The full list of questions is in \Cref{sec:tasks}.

\subsection{Experimental Setup}
\subsubsection{Experiment Coordination}
Two experimenters coordinated the study. The lead experimenter introduced participants to the concepts of human- and computer-controlled VHs. We avoided the technical terms of ``agent'' and ``avatar'' to prevent confusion. The lead experimenter also controlled the VH's interaction with the participant during the experiment, either directly via a head-mounted display (HMD) for the avatar phases or by triggering pre-recorded responses for the agent phases. In addition, the lead experimenter managed all procedures for the second phase of the experiment. The second experimenter greeted the participant, introduced them to the headset, handled the pre-experiment setup, and assisted with all participant-side procedures during \textit{Phase I}.

\subsubsection{Apparatus}
In \textit{Phase I}, both the participant and the experimenter used Meta Quest Pro headsets with head, hand, eye, and facial expression tracking to drive the VHs. The participant-side headset was connected via a wired Quest Link connection to a desktop computer with an NVIDIA GeForce RTX 3090 Ti GPU. The lead-experimenter-side headset was connected via a wired Quest Link connection to a laptop equipped with an NVIDIA GeForce RTX 4060 Laptop GPU. The participant and lead experimenter were in separate rooms to prevent the participant from knowing whether they were interacting with an agent or an avatar. The participant and lead experimenter remained standing in fixed positions. The guardian boundary for both the participant and the lead experimenter was set to approximately 2\,m $\times$ 2\,m. There was a physical table on the experimenter's side aligned with the virtual table in the virtual environment so that the experimenter could act naturally and rest their hands on the table. In \textit{Phase II}, the same desktop used on the participant's side in \textit{Phase I} was used to replay the recorded interactions. We used OBS Studio (version 29.1.1) to record the participant side computer display, with the participant's first-person Game view and a third-person Scene view in Unity. The interaction audio was recorded simultaneously.

\subsubsection{Virtual Environment}
\begin{figure}
    \centering
    \begin{subfigure}[t]{0.48\linewidth}
        \centering
        \includegraphics[width=\linewidth]
        {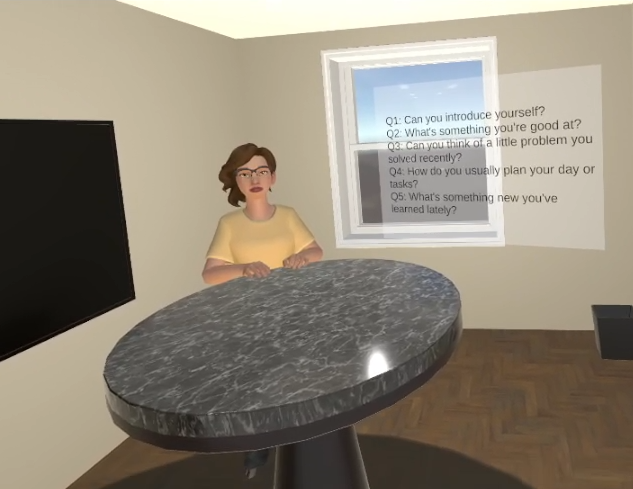}
        \caption{Participant view.}
        \label{fig:firstperson}
    \end{subfigure}
    \hfill
    \begin{subfigure}[t]{0.48\linewidth}
        \centering
        \includegraphics[width=\linewidth]
        {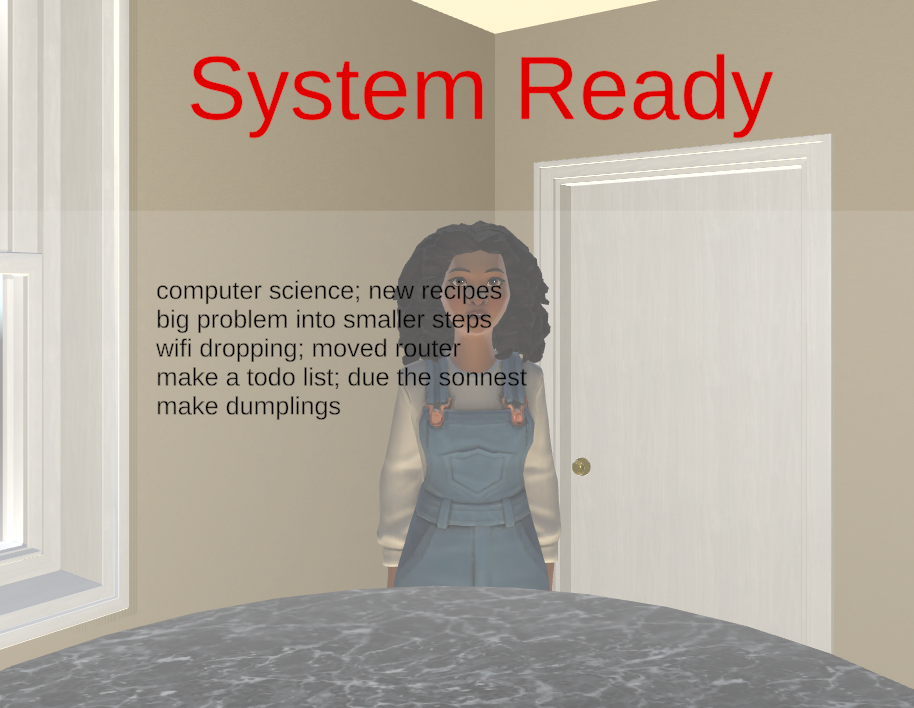}
        \caption{Lead experimenter view.}
        \label{fig:indicator}
    \end{subfigure}

    \caption{First-person interface perspectives.
    (a) The semi-transparent panel reminds participants of the
    predefined interview questions. The VH in the figure is controlled by the
    experimenter and can be either an agent or an avatar.
    (b) The red text indicates the control transition status, while
    the semi-transparent panel provides the scripted responses.
    The VH in the figure is an avatar controlled by the participant.}
    \label{fig:interface-perspectives}
\end{figure}

Two virtual scenes were used in \textit{Phase I} of the experiment. The preparation scene consisted of a floor and a mirror, allowing participants to familiarise themselves with their virtual bodies. The interaction scene was a virtual room designed to resemble a meeting room. Both the participant and the experimenter stood separated by a table of appropriate height. Given the limited field of view of the cameras on the Meta Quest headset, hands must be kept within the camera's view to ensure proper tracking. To ensure a natural appearance, the experimenter rested their hands on a physical table, which was adjusted to the same height as the virtual table. This made the hands of the experimenter-controlled VH appear to be resting naturally on the virtual table. \Cref{fig:firstperson} illustrates this environment and the appearance of the experimenter-controlled VH with hands on the table.


\subsection{VH Stimuli}
\subsubsection{VH Appearance}
\begin{figure}
    \centering
    \includegraphics[width=0.7\linewidth]{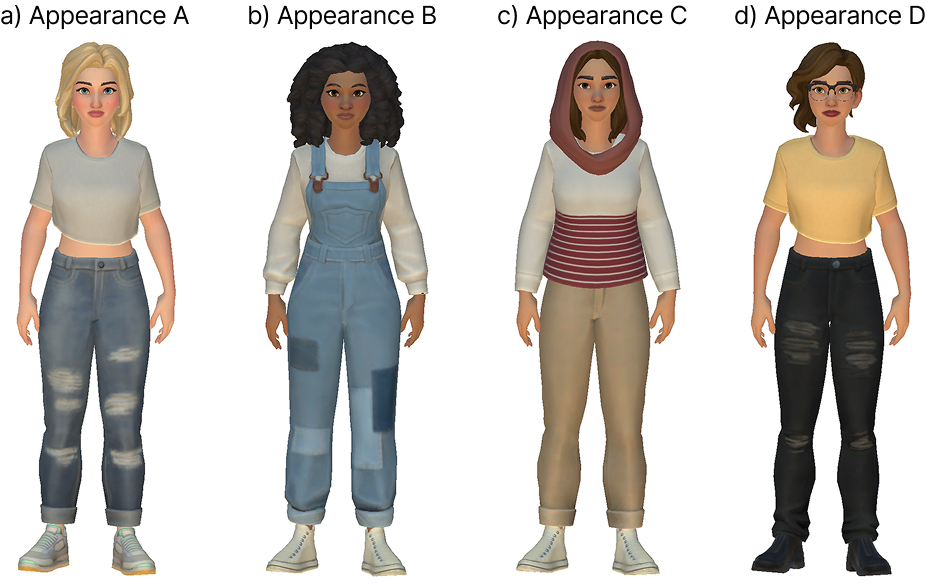}
    \caption{Different appearances of the VH used in the experiment.}
    \label{fig:appearance}
\end{figure}

We selected four VHs from the Meta Avatars SDK with different appearances from the built-in prefabs for interaction, as shown in \Cref{fig:appearance}. The Meta Avatars SDK provides full-body avatars with facial expressions and inverse kinematics to animate body parts using the available tracking data. We used version 35.0.0-pre.3. 
As the pre-recorded clips were voiced by the female lead experimenter, who also directly controlled the VHs during the experiment, all selected VHs were female in appearance to maintain consistency between voice and appearance and avoid introducing additional variability that might confound the results.  
The height of the camera rendering the user's view was adjusted to match the VH's head height at the beginning of each interaction session. 

\subsubsection{VH Responses}
We prepared four sets of pre-recorded responses for the agent phases. Each set contained similar answers, framed from the perspective of a university student, and each response was approximately one to two sentences in length. The goal was to ensure a consistent and non-distinctive impression across all sets. The specific responses can be found in the Supplementary Materials. Each set consisted of seven recordings: a looping ``standby'' motion, a non-verbal ``hi'' recording (a wave) for the initial greeting, and five recordings that included both verbal and non-verbal components as responses to each of the pre-defined interview questions. When we recorded these responses, we used a placeholder VH to represent the participant's future position, allowing us to naturally add body language and gestures to the recordings. Before each experiment, participants were asked to stand in approximately the same position as the placeholder. This ensured that the VH's body orientation and gaze were directed towards the participant during playback, making the VH appear to look at them.

The agent responses were delivered using a Wizard-of-Oz method, where the experimenter selected and played pre-recorded answers to simulate computer-controlled behaviour. This approach
ensured consistency across participants. 
The avatar was a VH directly controlled by the experimenter in real time. The experimenter followed the same scripts used for the pre-recorded responses.

\subsection{System}
We adapted the idea from Zhang et al. \cite{10937446} to develop a simple takeover system that allows control to transition smoothly between the agent and the avatar in either direction. The system for recording the VHs' responses was adapted from Brandstätter et al. \cite{brandstatter_record_2026}. Details of the networking and control transition and takeover processes are provided in the Supplementary Materials.  

\subsubsection{Virtual Humans}
The system incorporates four types of VHs. The Participant VH (PVH) is the avatar controlled by the participant in real time. The Experimenter VH (EVH) is the avatar controlled by the experimenter for interacting with the participant. The Interaction VH (IVH) is the VH whose motion and responses either replay pre-recorded animations or copy the motion of the EVH, depending on the current condition. The Placeholder VH (PHVH) is used on the experimenter's side to indicate the position and pose in the first frame of a replayed animation. It remains static to guide the experimenter before releasing control to the agent. Participants can see the PVH and IVH, while experimenters can see all four VHs.

\subsubsection{Interface} \label{sec: interface}

The lead experimenter's laptop interface includes seven buttons in the Unity Inspector: six for pre-recorded answers (including a ``hi" and five interview responses) and one ``takeover" button. Pressing an answer button applies the corresponding key tracking points to the IVH and replays the pre-recorded motion on both sides. Pressing the ``takeover" button transitions control from the experimenter's movements, tracked using the headset and hand tracking, to pre-recorded motions, or vice versa.

An indicator on the lead experimenter's side, shown in \Cref{fig:indicator}, indicates the status of the control transition. The text changes from ``System Ready" to ``Takeover in progress..." after the takeover button is clicked. Once the takeover process is complete, the text changes to ``Takeover Start''. A semi-transparent panel displays keywords from the scripted responses as reminders while the experimenter has full control of the VH, as shown in \Cref{fig:indicator}. Only the lead experimenter can see this panel. 


On the participant's side, a panel displays the predefined interview questions that participants should ask the VH during the task, as shown in \Cref{fig:firstperson}. Only the participant can see this panel.

\subsection{Procedure}

\textbf{Preparation}
Upon arrival, participants reviewed the participant information sheet and signed a consent form. They then completed a pre-experiment questionnaire (see \Cref{sec:Measure} for details). After that, the second experimenter assisted the participant in wearing the VR headset and explained the concept of hand-tracking technology.

Participants were guided into a virtual environment with a mirror in the preparation scene. They were instructed to make eye movements, facial expressions, and hand gestures to familiarize themselves with controlling a VH. This familiarization stage aimed to provide an understanding of what an avatar (human-controlled VH) looks like.

Afterward, the participant removed the headset and connected via a Microsoft Teams video call with the lead experimenter, who was located in a separate room. The lead experimenter explained the concepts of human-controlled and computer-controlled VHs and outlined the procedure for the first phase of the experiment. The lead experimenter then introduced the task that participants would perform during the interaction, and a practice session was conducted to ensure that participants understood the constraint of asking only the pre-defined questions without follow-ups. During this practice, the lead experimenter provided a one-sentence response to each question. These responses differed from those used in the main sessions but were kept consistent across all participants.

The second experimenter assisted the participant in putting the headset on for eye-tracking calibration. Once calibration was complete and the second experimenter confirmed with the lead experimenter that they were ready, the first phase of the experiment began. 

\textbf{Phase I} \label{sec: procedure phase i}
Upon entering the virtual meeting room, the participant's view was obscured by a blue curtain. During this time, the second experimenter started the recording, confirmed the proper functioning of eye and facial tracking on both sides and guided the participant to a designated standing position marked by two squares on the floor for foot placement. The session then started, and once it concluded, the participant removed the headset and completed the post-interaction questionnaire (see \Cref{sec:Measure} for details). This process was repeated four times until all conditions were covered.

\textbf{Phase II}
The lead experimenter then met the participant in person to begin the second phase of the experiment. Participants were informed that control might switch between a human and a computer during an interaction, although a switch would not necessarily occur. They then reviewed videos of their previous interactions while continuously adjusting a slider, with one end labelled ``Real Human'' (avatar) and the other labelled ``Computer Program'' (agent). The interface included both a third-person Unity scene-view recording of the VH they interacted with and a first-person game-view perspective, as shown in the Supplementary Materials. Participants were instructed to observe the video closely and provide reasons for any changes they made to the slider. 
After each clip, they were asked to explain in detail. 

After completing \textit{Phase II}, participants were informed how the VH had actually been controlled during each interaction. Following this disclosure, participants were asked whether they found conditions surprising and whether they had noticed changes in control.

\subsection{Measures} \label{sec:Measure}
Prior to the experiment, participants completed a pre-experiment questionnaire to provide basic demographic information on gender, age and VR experience. In addition, we investigated participants' perceived agency by collecting a combination of subjective (quantitative and qualitative) and objective (quantitative) data throughout the experiment. For complete questionnaires, see the Supplementary Materials. As the objective data are exploratory and do not contain any significant results, we include them in the Supplementary Materials.





\begin{figure*}
    \centering
    \includegraphics[width=0.8\textwidth]{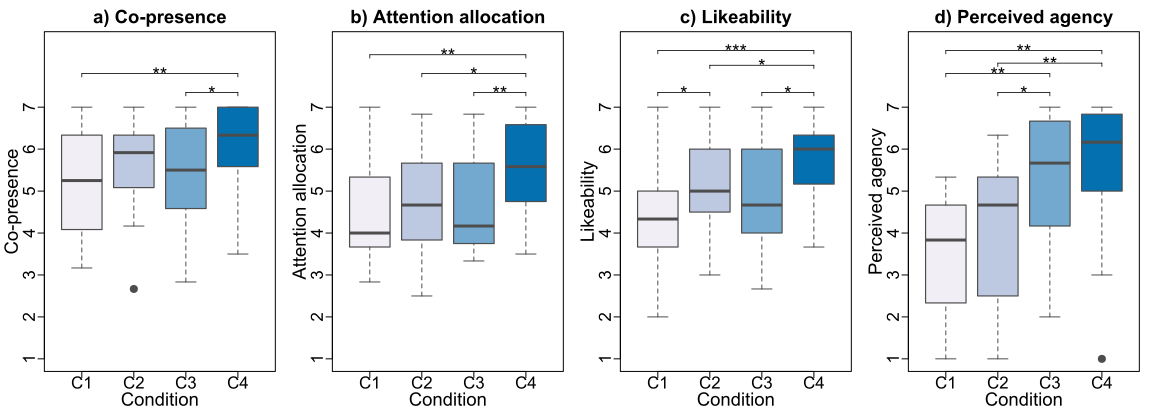}
    \caption{Post-interaction questionnaire scores across C1, C2, C3, and C4: (a) co-presence, (b) attention allocation, (c) likeability, and (d) perceived agency. Brackets indicate significant pairwise comparisons ($*p<.05$, $**p<.01$, $***p<.001$).}
    \label{fig:post-interaction-results}
\vspace{-6pt}
\end{figure*}

\subsubsection{Questionnaire Measures} 
After each session in the first phase of the experiment, participants completed a post-interaction questionnaire to assess social presence, likeability, perceived agency, and memory recall. As memory recall is exploratory and the result is not significant, we include it in the Supplementary Material.

\textbf{Social Presence:} We measured social presence using the questionnaire proposed by Harms et al. \cite{harms_internal_2004}, selecting the two subscales most relevant to our study: co-presence (6 items) and attention allocation (6 items). Each item was rated on a 7-point Likert scale. For co-presence, we averaged the scores to obtain a subscale score ranging from one to seven. For attention allocation, the four negatively worded items were reverse-scored before the scores were averaged to produce a subscale score ranging from one to seven.


\textbf{Likeability:} We used a custom three-item questionnaire to measure participants' likeability toward the VHs they interacted with. Items assessed participants' general impressions, including whether they liked the VH, enjoyed the interaction, or would like to meet the VH again. Responses were recorded on a 7-point Likert scale (1 = strongly disagree / not at all; 7 = strongly agree / very much). The three items were averaged to produce a likeability score ranging from one to seven.


\textbf{Perceived Agency:} To directly measure perceived agency, we used a three-item questionnaire assessing the extent to which the interviewee's overall behaviour, verbal behaviour, and non-verbal behaviour felt human- or computer-controlled. Participants rated each item on a 7-point Likert scale, with one end labelled ``Computer Program'' (coded as 1) and the other labelled ``Real Human'' (coded as 7). Scores were averaged to produce a perceived agency score ranging from one to seven.



\subsubsection{Slider and Interview Measures} \label{sec: subjective slider method}
In the second phase, as participants reviewed recordings of their interactions, we logged their continuous slider values and interviewed them about the reasons for their slider adjustments. Participants were asked to briefly state their reasons while adjusting the slider and to provide a more detailed explanation after watching the recording of each session. The slider was coded on a continuous scale from zero (Computer Program) to one (Real Human). The default slider value was set to the midpoint (0.5) at the beginning of each session. 

The start time of each interaction was defined as the moment at which the blue curtain blocking the participants' view while they were adjusting their position, as explained in \Cref{sec: procedure phase i}, was removed. Data from all interactions were divided into \emph{Before} and \emph{After} segments. For the consistent conditions C1 and C4, the split was set at the moment the participant began asking the third question. For the dynamic conditions C2 and C3, the split was set at the exact moment the takeover event was received on the participant side following the experimenter's button press, which occurred at approximately the time the participant began asking the third question.


For each segment, we calculated the time-weighted mean slider value. Specifically, we multiplied each slider value by the duration until the next recorded change (or the end of the segment), summed the resulting values, and divided the sum by the total duration of the segment. For both Before and After segments, the range of the time-weighted mean slider value is from zero to one.




\subsection{Hypothesis}
The four conditions varied in their initial controller, disclosed agency, and whether a control transition occurred. We therefore expected participants' responses to differ across conditions in:
\begin{itemize}
\item H1: Subjective measures including co-presence, attention allocation, likeability, and perceived agency.
\item H2: Time-weighted mean slider values in the Before and After segments.
\end{itemize}

%% file: 4.Result.tex
\section{Results}
We report the subjective measures including data from the self-reported questionnaire, slider, and interview. For all quantitative measures, we used Friedman tests to examine differences across the four experimental conditions. When a Friedman test revealed a significant effect, we performed pairwise post hoc comparisons using Wilcoxon signed-rank tests with Holm corrections for multiple comparisons. Adjusted $p$-values are reported below. In this section, we provide full details for statistically significant results, while the complete results are available in the Supplementary Materials.







\subsection{Questionnaire Measures}
The results from the post-interaction questionnaire are presented below and plotted in \Cref{fig:post-interaction-results}.

\textbf{Social Presence}
We identified that the co-presence score was significantly different across conditions, $\chi^2(3)=17.036,\ p<.001$. Post hoc analyses revealed significant differences between C1 and C4, $p =0.003$; and C3 and C4, $p = 0.013$.

For the attention allocation score, a significant difference was found across conditions, $\chi^2(3)=18.026,\ p<.001$. Post hoc analyses revealed significant differences between C1 and C4, $p =0.001$; C2 and C4, $p = 0.029$; and C3 and C4, $p = 0.009$.

\textbf{Likeability}
 The likeability score was significantly different across the four conditions, $\chi^2(3)=18.435,\ p<.001$. Post hoc analyses revealed significant differences between C1 and C2, $p = 0.023$; C1 and C4, $p < 0.001$; C2 and C4, $p = 0.023$; and C3 and C4, $p = 0.017$.

\textbf{Perceived Agency}
For perceived agency, a significant difference was found across the four conditions, $\chi^2(3)=30.120,\ p<.001$. Post hoc analyses revealed significant differences between C1 and C3, $p =0.004$; C1 and C4, $p =0.004$; C2 and C3, $p = 0.038$; C2 and C4, $p = 0.006$.


\subsection{Slider}


Slider values were recorded while participants watched recordings of their previous interactions in \textit{Phase II}. The data were divided into Before and After segments as described in \Cref{sec: subjective slider method}.

\begin{figure}[ht]
    \centering
    \includegraphics[width=0.6\linewidth]{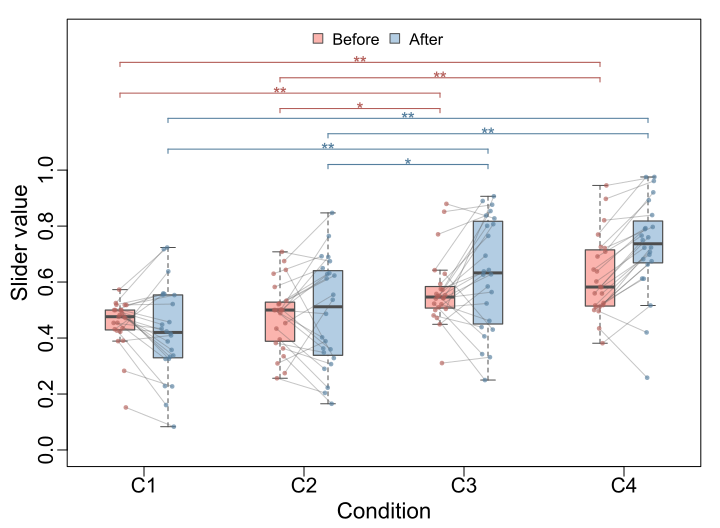}
    \caption{Time-weighted mean slider values in the Before and After segments across C1, C2, C3, and C4. Brackets indicate significant pairwise comparisons ($*p<.05$, $**p<.01$).}
    \label{fig:slider}
\end{figure}

The time-weighted mean slider values in the Before segment differed significantly across conditions, $\chi^2(3)=25.190,\ p<0.001$, as shown in \Cref{fig:slider}. Post hoc analyses revealed significant differences between C1 and C3, $p =0.004$; C1 and C4, $p = 0.002$; C2 and C3, $p = 0.021$; C2 and C4, $p =0.007$.

The time-weighted mean slider values in the After segment also differed significantly across conditions, $\chi^2(3)=26.150,\ p<0.001$, as shown in \Cref{fig:slider}. Post hoc analyses revealed significant differences between C1 and C3, $p =0.006$; C1 and C4, $p = 0.001$; C2 and C3, $p = 0.044$; C2 and C4, $p =0.002$.

To further investigate whether slider ratings changed between segments within each condition, we conducted paired Wilcoxon signed-rank tests comparing the time-weighted mean slider values in the Before and After segments, with Holm correction for four comparisons. Only C4 showed a significant increase from Before to After, $V=276,\ p=0.001$. No significant changes were found in C1, C2, or C3.





\subsection{Interview}

The interviews provided insights into how participants interpreted the VHs' behaviour as evidence of whether the VHs were human- or computer-controlled. Their explanations referred both to particular behaviours and to how those behaviours related to the content of an answer or occurred in response to participants' own actions. We first report four frequently mentioned aspects: body movement and gesture, facial expression, gaze and eye behaviour, and response timing. We then report how prior information about the controller shaped some participants' interpretations of the behaviours they observed. Finally, we describe how some participants perceived different forms of control during and between answers.


\textbf{Body Movement and Gesture.}
All 24 participants referred to at least one aspect of body movement or gesture as a factor in their agency judgements. Participants highlighted five dimensions, including amount, amplitude, perceived quality, repetition, and correspondence with speech. For additional examples and quotations from participants, see the Supplementary Materials.

More frequent or continuous movement supported human control for 11 participants, whereas limited or absent movement supported computer control for 13. For example, P5 described a VH as ``static like a doll'' while P17 noted that another kept its hands ``always still on the desk''.

Participants used movement amplitude and quality as evidence of human or computer control. Small or subtle movements supported human control for 11 participants but computer control for four. Larger movements were associated with human control by four participants. P1 associated ``large motion'' with ``slightly more passion''. In contrast, two participants reported that excessive or abruptly enlarged movements could weaken the impression of human control. Four participants associated smooth, organic, flexible, or detailed movements with human control, although P19 described one VH as ``too smooth'' and therefore ``a bit like a computer''. Eight participants associated movement described as jittery, stiff, choreographed, pre-designed, robotic, synthetic, or strange with computer control. Here, ``pre-designed'' referred to how formulaic the gestures appeared; P10 described them as ``routine gestures that people in the interview will have''. Repetitive or patterned movements supported computer control for five participants.

Twelve participants used correspondence between gestures and speech, perceived emotion, facial expression, or eye behaviour as evidence of human control, whereas four used mismatches as evidence of computer control. However, synchrony alone did not necessarily indicate human control. P8, for example, described the gestures as ``extremely synced'' and therefore ``too good to be true''. 

The greeting wave was discussed by 12 participants. Ten participants used it as evidence of human control and five as evidence of computer control, with three participants using it in both directions. P6, for example, said that the wave made the VH seem ``a lot more engaged''. P13 instead perceived the greeting as unnatural and described it as ``just like a robot''.

\textbf{Facial Expression.}
Facial expressions or displayed emotion were used as evidence in agency judgements of 14 participants. Natural or clearly discernible expressions supported human control, whereas eight participants associated absent, weak, or unchanged expressions with computer control. Smiling was the most frequently identified facial expression, with seven participants interpreting it as evidence of human control. However, some smiles were perceived as ``synthetic'' (P14). P11 interpreted smiles that disappeared too quickly as unnatural and therefore as evidence of computer control.



\textbf{Gaze and Eye Behaviour.}
Thirteen participants agreed that maintaining eye contact was a reason for perceiving a VH as human-controlled. Conversely, a lack of eye contact was an indicator of an agent for 15 participants. For example, P1 noted that the VH ``remains stick towards one position without tiny movement to follow my head,'' which they interpreted as a sign of computer control. However, three participants considered not maintaining continuous eye contact to be more human-like. P8 pointed out that ``a little bit of eye avoiding'' felt more human, whereas a VH that ``looked at me all the time'' appeared ``not 100\% human.''

Natural eye movements were mentioned by 20 participants as a reason for perceiving the VH as human-controlled, while absent or unnatural eye movements were stated by 11 participants as a sign of computer control. P11 stated, ``rolling the eyes during the speech'' made the VH feel human-like, while P4 described the VH ``looking at the same direction all the time'' as computer-program-like.



\textbf{Response Timing.}
Response timing also influenced participants' judgements. Empty or unexplained delays before answers were interpreted as computer processing by 13 participants. P18 noted that ``Real human, they wouldn't take that long to think ... because this is just some simple question.''

However, 19 participants interpreted observable thinking behaviours while an answer was being formulated as evidence of human control. These behaviours included gaze shifts during thinking, mentioned by 17 participants; body movements, mentioned by 10; fillers or filled starts such as ``umm'', mentioned by four; and facial expressions, mentioned by two.

Two participants did not mention thinking behaviours during the pause, but they still believed a short pause made the VH seem more human. P14 stated that ``humans don't ... have the answer directly in their head'' and are ``not like a robot memorised all the answers.'' Conversely, three participants interpreted an immediate or fluent response as evidence of computer control because it appeared to lack a thinking process.

One participant also attended to the pattern of response timing across the interaction. P22 explained that ``in real life scenarios, we don't hesitate all the time with the same time period.'' They believed that repeated hesitations of the same length supported computer control, whereas randomly occurring hesitation supported human control.


\textbf{Prior Information.}
Five participants mentioned that the prior disclosure of agency could influence their judgements in different ways. P20 acknowledged, ``I just dragged to the left because I was told it's controlled by a computer.'' P12 reported looking specifically for possible lip-synchronisation problems after being told in \textit{Phase I} that the VH was computer-controlled. P1 made the assumption that the controller would remain the same throughout a run, explaining that ``... in some part it feels like it's a real human, but according to the him that there is only one, like it should be either a computer program or a human.'' P2 explained that believing the VH was controlled by a real human may have led them to interpret the response delays as thinking. P13 also mentioned that the provided information influenced their answer but did not elaborate further. However, one participant (P18) also mentioned trying to ``just ignore'' what the experimenter told them.

\textbf{Behaviour During and Between Answers}
Three participants interpreted the VH as human-controlled while it was answering but as computer-controlled between answers. For example, P15 said that the VH's hand and facial movements appeared natural during answers, whereas its behaviour between answers felt ``static and automatic and computer-controlled''.




%% file: 5.Discussion.tex
\section{Discussion}
In this section, we interpret the quantitative and qualitative findings concerning perceived agency and participants' social responses to the virtual human. We then consider the limitations of the study and directions for future work.


\subsection{Post-interaction Perceived Agency} \label{sec: post-interaction discussion}
The post-interaction questionnaire revealed a significant effect of condition on perceived agency, with median scores increasing from C1 to C4. This overall condition effect is consistent with H1, which predicted differences in perceived agency.
The post-hoc analysis suggested that the conditions could be grouped according to the agency established at the beginning of the interaction: C3 and C4 (both human-controlled during the Before segment and disclosed as human-controlled) received significantly higher perceived agency scores than C1 and C2 (both computer-controlled during the Before segment and disclosed as computer-controlled). In other words, the initial-agency group that began with human control received higher perceived agency scores than the initial-agency group that began with computer control.

Within each of the initial-agency groups, C2, which transitioned to human control, was rated higher than the consistently computer-controlled C1, whereas C3, which transitioned to computer control, was rated lower than the consistently human-controlled C4. Although neither within-group comparison was significant, both descriptive differences followed the direction expected if the controller present later in the interaction had a smaller influence on perceived agency, shifting ratings towards the corresponding form of control. Taken together, the results suggest that perceived agency was primarily structured by the agency established at the beginning of the interaction, with a smaller and statistically non-significant contribution from the controller present later in the interaction.

The difference in perceived agency between the two initial-agency groups may have resulted from the combined influence of the agency disclosure and the behaviour observed during the Before segment, which together established participants' first impressions of who was controlling the VH. The disclosure provided a prior belief and expectation about the controller, while the behaviour of the aligned initial controller provided congruent perceptual evidence. Together, these two sources may have reinforced a relatively stable initial belief about who was controlling the VH, making that belief less likely to be revised by inconsistent cues encountered during the After segment. This interpretation is consistent with prior work showing that expectations can have a stronger influence when behavioural evidence is ambiguous \cite{epley_when_2005}, and that strong prior impressions can persist when inconsistent information can be explained in alternative ways \cite{kim_psychology_2020}.

\subsection{Dynamics of Perceived Agency} \label{sec: discuss behavioural realism}
The slider ratings showed a pattern consistent with the post-interaction perceived-agency results. Mean slider values increased from C1 to C4 in both the Before and After segments, supporting H2, which predicted condition differences in time-weighted slider judgements across the two segments. Across both segments, all comparisons between conditions disclosed as computer-controlled (C1 and C2) and those disclosed as human-controlled (C3 and C4) were significant, whereas the difference between C1 and C2 and the difference between C3 and C4 were not significant. Importantly, in the After segment, conditions that shared the same disclosed agency and initial controller did not differ significantly, whereas conditions that shared the same controller during that segment did. This pattern suggests that participants' later judgements remained anchored to the agency established during the initial part of the interaction.

The interviews showed the same type of behavioural cue could support slider movements towards either human or computer control. For example, smooth movement could support human control, whereas movement perceived as ``too smooth'' could support computer control. These contrasting interpretations may have reflected subtle differences in the timing and naturalness of the observed behaviour. They may also have reflected participants interpreting ambiguous cues in ways consistent with their existing agency beliefs. Five participants reported that the initial disclosure influenced their judgements. Prior belief may also have influenced the remaining participants' slider judgements without being explicitly reported, although the interviews cannot establish this because they captured only participants' stated reasons. As discussed in \Cref{sec: post-interaction discussion}, this possibility is consistent with prior work on the influence of expectations under ambiguous behavioural evidence \cite{epley_when_2005}.

Only C4 showed a significant increase from the Before to the After segment, with slider ratings moving further towards the human side. In comparison, C1 showed only a small and non-significant decrease towards the computer side, while no significant Before and After changes were found in C2 or C3. This asymmetry may suggest that participants were more sensitive to cues indicating human control than to cues indicating computer control. Continued human control in C4 made participants more certain that the VH was controlled by a person, whereas continued computer control did not produce an equally clear slider movement towards the computer side. The slider values in the conditions disclosed as computer-controlled (C1 and C2) remained closer to the midpoint, while those in the conditions disclosed as human-controlled (C3 and C4) showed a clearer tendency towards the human side. It could be that participants did not perceive the later interaction in C3 as becoming more computer-like following the transition. Another possibility is that the absence of a clear shift towards the computer side was related to the way the computer-controlled VH was simulated.

The Wizard-of-Oz setup used to simulate the computer-controlled VH may explain this pattern. The computer-controlled VH was simulated by replaying speech and movements recorded from an actual human performance. This design enabled us to simulate a computer-controlled VH with human-like behaviour while keeping its responses consistent across participants. Consequently, the behavioural distinction between the simulated agent and live human control may have been relatively subtle. This interpretation is consistent with previous findings that human-like non-verbal behaviour can increase humanness attribution \cite{pfeiffer_non-verbal_2011}, and that behavioural realism can be more influential than disclosed agency alone \cite{von_der_putten_it_2010,guadagno_virtual_2007}. Thus, the slider values may have shifted towards the human side. However, the significant differences between conditions disclosed as computer-controlled (C1 and C2) and those disclosed as human-controlled (C3 and C4) show that participants still distinguished the two sets of conditions. The limited movement towards the computer side may therefore reflect a less pronounced attribution of computer control rather than participants clearly identifying the pre-recorded behaviour as agent-controlled.

The interviews also suggest that behavioural realism played a role in perceived agency. Behaviour appeared to be natural, contextually appropriate, and responsive supported human control, whereas stillness, repetition, mismatches among behaviours, and unexplained delays supported computer control. Behavioural realism was therefore one source of evidence used to judge who was currently controlling the VH. However, because real-time human control was expected to provide greater behavioural realism, a larger proportion of human-controlled interaction would also be expected to result in slider ratings further towards human control. The ratings did not fully follow this pattern. C3 received more human-oriented slider ratings than C2 despite containing a shorter human-controlled proportion. This suggests that behavioural realism contributed to the slider values but did not outweigh the agency belief established through the disclosure and initial interaction.




\subsection{Social Presence}
Both subscales of social presence, co-presence and attention allocation, differed significantly across conditions, supporting H1. 
Scores on both subscales increased as the proportion of the interaction under human control increased, following the order C1, C3, C2, and C4.
Specifically, C1 contained no human-controlled segment, C3 was human-controlled during the first two questions, C2 during the final three questions, and C4 throughout all five questions. The significant difference between C1 (always agent-controlled) and C4 (always human-controlled) is consistent with meta-analytic evidence showing that perceived avatars elicit greater social presence than perceived agents \cite{felnhofer_virtual_2024}. However, the social-presence subscales for C2 and C3 did not follow the ordering observed for perceived agency, suggesting that social presence may also have been shaped by the behaviour experienced during the interaction.

One reason for this ordering may be that a greater proportion of the interaction under human control created a stronger overall impression of behavioural realism. Although the human controller followed the same response scripts as the pre-recorded agent, real-time human control could still introduce subtle temporal and behavioural differences. These could include response timing adapted to the participant's speech, spontaneous pauses and gaze shifts while preparing an answer, continuous postural and facial adjustments, eye contact with the participant, and responses to the participant's movements. Such behaviours could increase the behavioural realism of the VH and make it appear more responsive and involved, thereby increasing both the participant's sense of co-presence and the attention allocated to the VH. This aligns with prior work showing that behavioural realism contributes to social presence and social responses towards virtual humans \cite{kyrlitsias_social_2022,guadagno_virtual_2007,von_der_putten_it_2010}.

The ordering of the social-presence subscales may also reflect whether human control occurred near the end of the interaction and therefore closer to the administration of the post-interaction questionnaire. The conditions involving a greater proportion of interaction under human control were also those in which human control occurred later in the interaction. Previous research has shown that the sequencing of interaction experiences, particularly their most recent moments, can influence retrospective evaluations \cite{cockburn_effects_2017}. The higher descriptive social-presence scores in C2 than in C3 could therefore reflect longer human exposure, more recent human exposure, or a combination of both.

\subsection{Likeability}
Likeability also differed significantly across conditions, supporting H1. Likeability scores followed the same ordering as the two social-presence subscales, increasing from C1 to C3, C2, and C4, rather than the ordering observed for perceived agency. This correspondence is consistent with prior work showing that greater behavioural realism is associated with greater social presence and more positive interpersonal evaluations, including greater interpersonal attraction \cite{herrera_effect_2018}. Notably, likeability was significantly higher in C2 than in C1 and in C4 than in C3. Within each pair, the conditions shared the same disclosed agency and initial controller but differed in the controller during the later part of the interaction. Participants therefore evaluated the VH more favourably when the later part of the interaction was human-controlled. This pattern suggests that participants' evaluations of the VH were more closely aligned with their ratings of social presence than with their perceived agency.

\subsection{Social Response Models} \label{sec: model discuss}
The two social presence subscales and likeability differed significantly across conditions that showed differences in perceived agency. C1 (always agent) with the lowest perceived agency received the lowest social response scores, whereas C4 (always avatar) with the highest perceived agency received the highest social response scores. However, the ordering of C2 and C3 followed the proportion and timing of real-time human control rather than the ordering of their perceived-agency scores. Taken together, this pattern suggests that participants' social responses were influenced both by the perceived agency established at the beginning of the interaction and by the behavioural realism experienced throughout it. Compared with the Media Equation Concept described in \Cref{sec: related work social response models}, which proposes that social reactions arise independently of perceived agency, our findings are more consistent with the Threshold Model of Social Influence, which predicts that perceived agency and behavioural realism jointly shape social influence.

\subsection{Limitations}

The agent phases used a Wizard-of-Oz setup in which pre-recorded human speech and movements were presented as computer-controlled behaviour. This method allowed us to simulate an advanced agent capable of producing highly human-like responses for future hybrid-control applications without the need to develop a generative model trained to clone the experimenter's voice and movements, which is beyond the scope of this study. At the time of the study, the differences between the available generated behaviour and the experimenter's live performance would likely have made the controller transition easy to detect. However, this design also meant that the study compared real-time and pre-recorded human performances rather than real-time human and fully computer-generated behaviour. Nevertheless, the findings provide insight into how perceived agency and social responses are shaped by controller transitions in hybrid systems where agent behaviour is designed to closely approximate live human performance.

The human-controlled interactions could not be standardised across participants. Although the human controller followed the same response scripts and behavioural guidelines, subtle variations in response timing, gaze, facial expressions, posture, and reactions to individual participants were unavoidable. However, this variability was also integral to the comparison because the human controller's ability to adapt naturally to each participant was one way in which real-time human control differed from pre-recorded agent behaviour and may have increased behavioural realism \cite{kyrlitsias_social_2022}. Therefore, this variability does not prevent the findings from being interpreted as differences between the two control modes.

Another limitation concerns the appearance of the participant's self-avatar. All participants embodied the same self-avatar regardless of their gender, and its appearance was identical to one of the four VH appearances used for the interaction partners (Appearance B). Encountering a VH with the same appearance as one's embodied avatar may have affected familiarity or evaluations of that VH. However, Appearance B was counterbalanced across the four conditions and the participant's self-avatar remained constant across sessions, meaning that these factors were not systematically confounded with condition.



\subsection{Future Work}
It remains unclear how firmly an agency belief becomes established at the beginning of an interaction, and whether this depends more on the prior belief created by agency disclosure or on the first impression formed through the initial interaction. Prior studies have manipulated agency beliefs by having a computer program control the VH in all conditions while presenting it as either human- or computer-controlled \cite{fox_avatars_2015,felnhofer_virtual_2024}. Future work could compare a fully generated agent with real-time human control to determine whether participants' agency judgements would remain anchored to their prior beliefs when the behavioural difference between the controllers is larger. However, would a human-control disclosure unsupported by the initial interaction produce an agency belief that is more easily revised by subsequent cues? Does disclosure or the initial interaction contribute more strongly to the robustness of the resulting belief? What makes such a belief more resistant to contradictory evidence or, conversely, more easily doubted?

The dynamics of perceived agency remain unclear as well. Future studies could systematically vary the duration of the post-transition interaction and the strength of the behavioural change. The open questions are how strong behavioural cues must be to challenge an established belief about who is controlling the VH, and how long participants must be exposed to these cues before that belief begins to change. Repeated agency measurements could help determine whether perceived agency is gradually influenced as evidence accumulates or changes only when the behavioural cues become sufficiently strong. Such investigations would extend research on the persistence of initial impressions \cite{kim_psychology_2020} and belief updating following unexpected social behaviour \cite{christian_belief_2024}.

However, it could be difficult to measure perceived agency repeatedly throughout the interaction. Interrupting the interaction to obtain repeated ratings could disrupt its flow and cause a break in presence in VR \cite{putze_breaking_2020}. Repeatedly asking about agency could also expose the purpose of the experiment, directing participants' attention towards behavioural cues that they might not otherwise monitor during a natural interaction. Future work could therefore use objective behavioural measures alongside subjective ratings as indirect indicators of how participants respond to the perceived controller. For example, prior work found that participants regulated their interpersonal distance differently depending on whether a VH was presented as an agent or an avatar \cite{bailenson_interpersonal_2003}. Nevertheless, a recent meta-analysis found no reliable effect of agency on behavioural outcomes in immersive VR \cite{felnhofer_virtual_2024}, consistent with the absence of significant objective behavioural differences in this study.






%% file: 6.Conclusion.tex
\section{Conclusion}
In conclusion, we investigated how perceived agency develops when control of a VH transitions between a computer-controlled agent and a human-controlled avatar during an ongoing VR interaction. We conducted a two-phase within-subject study with 24 participants. In the first phase, participants interacted with VHs under four agency conditions without being informed that control could transition. In the second phase, participants reviewed recordings of the interactions, continuously rated perceived agency using a slider, and explained their judgements. The results revealed that perceived agency remained anchored to the agency disclosure and initial interaction, with no evidence of updating following the control transitions, whereas social presence and likeability appeared more responsive to the proportion of the interaction under human control. The interview findings suggested that participants judged agency based on behavioural cues and their prior beliefs about agency. These findings provide insights into how perceived agency and social evaluations develop during interactions involving dynamic control and may inform the design of future hybrid systems that transition between increasingly realistic AI agents or AI clones, and human operators.

%% file: supplement.tex
\section{System}

\subsection{Networking}
Our system uses Ubiq version 1.0.0-pre.8 \cite{fristonUbiqSystemBuild2021}, a lightweight Unity-based networking package, to synchronise users' locations, gestures, voice communication, message passing, and object spawning. Both the participant and experimenter sides ran the client application built with the Ubiq package and joined the same Ubiq room. Key tracking points of the PVH and EVH are shared through real-time messages. While the participant side does not render the EVH, it still receives the EVH's key-tracking-point data. The IVH is treated as a networked object, so its key tracking points are also shared across the network and synchronised on both sides. The PHVH remains local to the experimenter's side.

\subsection{Control Transition} \label{sec: control transition}
In the avatar phase, the IVH on the participant's side directly copies the motions of the EVH. In the agent phase, the IVH on the participant's side replays and loops a pre-recorded standby motion. When the experimenter presses an answer button, a message is sent to the participant's side, causing the IVH to replay the corresponding animation. Once the replay concludes, the IVH returns to the standby loop.

For C2, before the experimenter presses the takeover button, they align their posture with the IVH's current pose. Upon pressing the button, the IVH on the participant's side stops replaying and smoothly blends into the experimenter's current key tracking points using linear interpolation. Once blending is complete, the avatar phase begins. 

For C3, the system begins in the avatar phase. The experimenter's side displays the PHVH, which remains static at the first frame of the pre-recorded standby motion. Before pressing the takeover button, the experimenter aligns their posture with the PHVH's pose.
Once the button is pressed, the PHVH disappears and the IVH on the participant's side stops copying the EVH and smoothly blends into the first frame of the standby animation. When blending is complete, the agent phase begins.

The transition blend duration was 0.444 s on average (SD = 0.018 s).

\newpage
\section{Phase II Review Interface}

\begin{figure}[h]
    \centering
    \includegraphics[width=\linewidth]{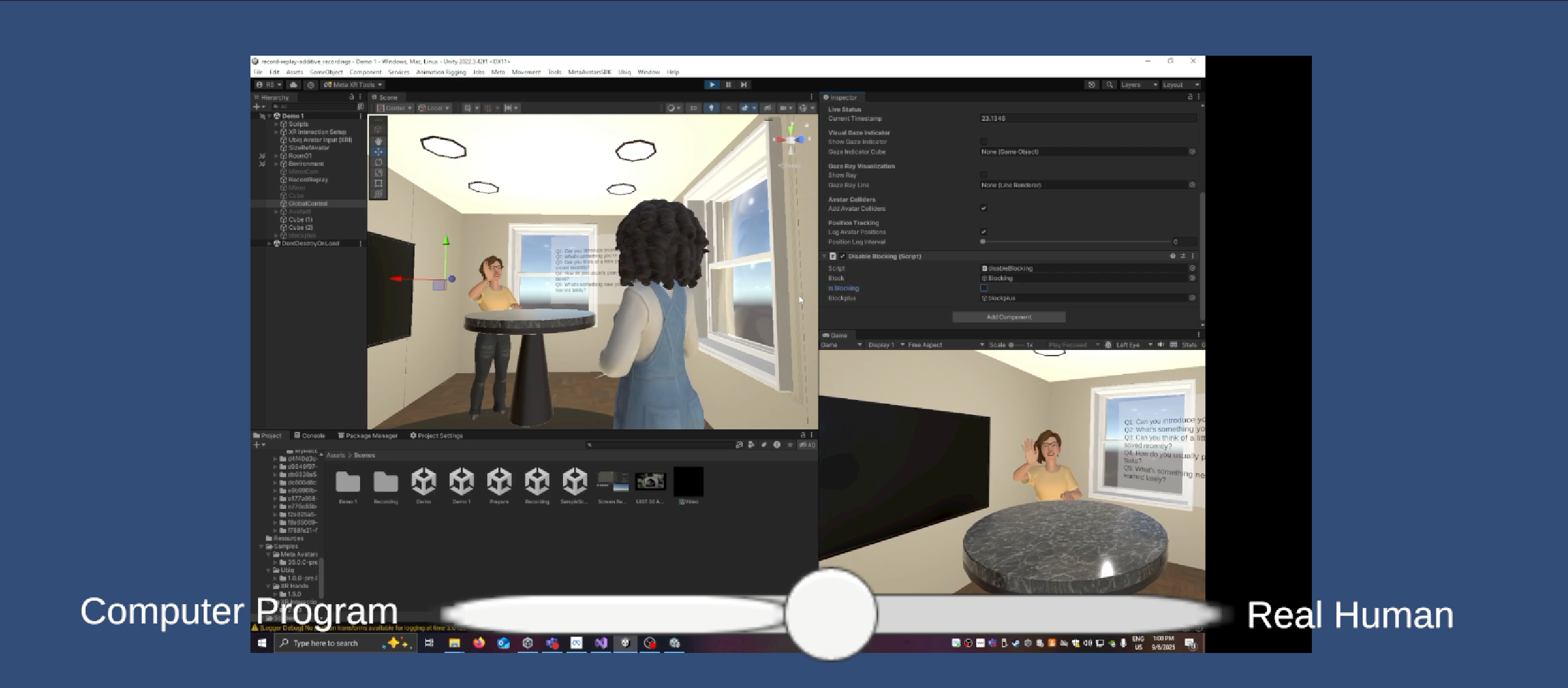}
    \caption{A screenshot showing the interface included both a third-person Unity scene-view recording of the VH they interacted with and a first-person game-view perspective and the slider used in Phase II.}
    \label{fig:placeholder}
\end{figure}

\newpage
\section{Results}
For all quantitative measures, we used Friedman tests to examine differences across the four experimental conditions. When a Friedman test revealed a significant effect, we performed pairwise post hoc comparisons using Wilcoxon signed-rank tests with Holm corrections for multiple comparisons. Adjusted p-values are reported below.

\subsection{Objective Measures}
Throughout \textit{Phase I} of the experiment, we logged objective data from the participant, including gaze and head position. The start time of each interaction was defined as the moment at which the blue curtain blocking the participants' view while they were adjusting their position, was removed. Data from all interactions were divided into \emph{Before} and \emph{After} segments. For the consistent conditions C1 and C4, the split was set at the moment the participant began asking the third question. For the dynamic conditions C2 and C3, the split was set at the exact moment the takeover event was received on the participant side following the experimenter's button press, which occurred at approximately the time the participant began asking the third question. Results for objective measures are in \cref{tab:gaze-head-descriptives}.

\begin{table*}[h]
\centering
\caption{Descriptive statistics for gaze and head displacement across conditions and segments. Values are reported as median (IQR).}
\label{tab:gaze-head-descriptives}
\small
\begin{tabular}{lcccc}
\toprule
& \multicolumn{2}{c}{\emph{Before}}
& \multicolumn{2}{c}{\emph{After}} \\
\cmidrule(lr){2-3}\cmidrule(lr){4-5}
Condition
& Gaze at VH (\%)
& Head displacement (m)
& Gaze at VH (\%)
& Head displacement (m) \\
\midrule
C1
& 48.908 (67.755)
& 0.028 (0.045)
& 45.314 (65.433)
& 0.037 (0.043) \\

C2
& 62.660 (42.482)
& 0.046 (0.024)
& 53.294 (62.329)
& 0.059 (0.064) \\

C3
& 21.227 (55.306)
& 0.040 (0.036)
& 32.719 (62.469)
& 0.045 (0.054) \\

C4
& 45.653 (66.274)
& 0.044 (0.028)
& 47.079 (46.194)
& 0.054 (0.055) \\
\bottomrule
\end{tabular}
\end{table*}

\subsubsection{Gaze}
We logged participants' gaze targets in the virtual environment, including their own VH (head, hands, and body), the VH with which they interacted (head, hands, and body), the reminder panel displaying the predefined interview questions, and the surrounding environment. We measured the percentage of time participants gazed at the VH with which they interacted during the Before and After segments.

A quality check identified limitations in the gaze recordings. Of the 192 observations (24 participants $\times$ four conditions $\times$ two segments), 49 (25.5\%) had a recorded gaze-at-VH percentage of zero, corresponding to 25 of the 96 interactions. In 24 interactions, the recorded percentage was zero in both the \emph{Before} and \emph{After} segments; in one interaction, it was zero only in the \emph{Before} segment. Eight interactions showed clearly invalid tracking data, with the recorded eye position and gaze direction remaining fixed at $(0,0,0)$ and $(0,0,1)$, respectively, throughout the analysed interval.

The proportion of time spent looking at the partner VH did not differ significantly across conditions in either the Before segment, \(\chi^2(3)=4.441,\ p=.218\), or the After segment, \(\chi^2(3)=3.815,\ p=.282\).  

\subsubsection{Head Position}
 We logged the three-dimensional position of the head of each participant's avatar. The baseline was defined as the mean head position during the first second of each interaction. For every recorded frame, we calculated head displacement as the Euclidean distance between the current head position and this baseline position. We then calculated the time-weighted mean head displacement separately for the Before and After segments.

 No significant differences in time-weighted mean head displacement were found across conditions in either the Before segment, $\chi^2(3)=2.150,\ p=.542$, or the After segment, $\chi^2(3)=6.650,\ p=0.084$.


\subsection{Subjective Measures}
The median and IQR of all subjective measures are in \cref{tab:subjective-descriptive}.

\begin{table*}[h]
\centering
\caption{Descriptive statistics for the subjective measures across conditions. Values are reported as median (IQR), with $n=24$ per condition. Questionnaire scores ranged from 1 to 7, except memory recall, which ranged from 0 to 3. Slider values ranged from 0 to 1.}
\label{tab:subjective-descriptive}
\begin{tabular}{lcccc}
\toprule
Measure
& C1
& C2
& C3
& C4 \\
\midrule
\multicolumn{5}{l}{\textit{Questionnaire measures}} \\
Co-presence
& 5.250 (2.208)
& 5.917 (1.208)
& 5.500 (1.875)
& 6.333 (1.375) \\
Attention allocation
& 4.000 (1.583)
& 4.667 (1.833)
& 4.167 (1.875)
& 5.583 (1.750) \\
Likeability
& 4.333 (1.333)
& 5.000 (1.417)
& 4.667 (2.000)
& 6.000 (1.083) \\
Perceived agency
& 3.833 (2.333)
& 4.667 (2.750)
& 5.667 (2.250)
& 6.167 (1.750) \\
Memory recall
& 3.000 (1.000)
& 2.500 (1.000)
& 3.000 (1.000)
& 3.000 (1.000) \\
\addlinespace
\multicolumn{5}{l}{\textit{Time-weighted mean slider values}} \\
Before
& 0.476 (0.070)
& 0.500 (0.134)
& 0.547 (0.071)
& 0.582 (0.196) \\
After
& 0.420 (0.223)
& 0.512 (0.293)
& 0.633 (0.356)
& 0.737 (0.136) \\
\bottomrule
\end{tabular}
\end{table*}

\subsubsection{Co-presence}
Post hoc analyses revealed non-significant differences between C1 and C2, $p =0.388$; C1 and C3, $p = 0.560$; C2 and C3, $p = 0.560$; C2 and C4, $p = 0.057$.

\subsubsection{Attention Allocation}
Post hoc analyses revealed non-significant differences between C1 and C2, $p =0.419$; C1 and C3, $p = 0.930$; C2 and C3, $p = 0.930$.

\subsubsection{Likeability}
Post hoc analyses revealed non-significant differences between C1 and C3, $p = 0.681$; C2 and C3, $p = 0.681$.

\subsubsection{Perceived Agency}
Post hoc analyses revealed non-significant differences between C1 and C2, $p =0.163$; C3 and C4, $p = 0.163$.

\subsubsection{Memory Recall} 
We investigated participants’ memory of the interactions by asking them to recall specific details, including: ``What subject is the interviewee studying?'', ``What did the interviewee say they recently fixed or solved?'', and ``What did the interviewee recently learn to make or do?''. Each question had four choices plus an ``I don’t remember'' option. A correct answer was coded as one, while all other responses were coded as zero. The scores were summed to produce a total memory recall score ranging from zero to three. 

No significant difference was found for the memory recall score, $\chi^2(3)=0.108,\ p=0.991$.

\subsection{Slider}
Post hoc analyses of the time-weighted mean slider values in the Before segment revealed non-significant differences between C1 and C2, $p =0.745$; C3 and C4, $p =0.213$.

Post hoc analyses of the time-weighted mean slider values in the After segment revealed non-significant differences between C1 and C2, $p =0.259$; C3 and C4, $p =0.054$.

To further investigate whether slider ratings changed between segments within each condition, we conducted paired Wilcoxon signed-rank tests comparing the time-weighted mean slider values in the Before and After segments, with Holm correction for four comparisons. No significant difference was found in C1 between Before and After, ($V=103, p=0.368$). No significant difference was found in C2 between Before and After, ($V=164, p=0.700$). No significant difference was found in C3 between Before and After ($V=213, p=0.222$).

\subsection{Interview}
Below we present more detailed results for the Body Movement and Gesture section in the main paper.

The overall amount of movement shaped participants' judgements. More frequent or continuous movement was used as evidence of human control by 11 participants. For example, P18 became ``more confirmed that it was controlled by a human'' as the VH showed ``more body language''. P21 similarly emphasised that the VH was ``always moving''. Conversely, limited or absent movement supported computer control for 13 participants. Participants described VHs as ``static like a doll'' (P5) or keeping their hands ``always still on the desk'' (P17).

Participants used movement amplitude to support both human- and computer-control judgements. Small or subtle movements were associated with human control by 11 participants; these included movements of the head, hands, fingers, arms, shoulders, and whole body. However, four participants associated small movements with computer control. P1 illustrated this contrast: ``tiny motion'' made one VH appear ``more like a human'', whereas another appeared computer-controlled because it showed ``no body movement or just tiny body movement all the time''. P14 similarly interpreted a ``very small hand movement'' as appearing to ``mimic a human rather than [result from] a human controlling it''. Larger movements were associated with human control by four participants. P1 associated ``large motion'' with ``slightly more passion''. In contrast, movements that appeared excessive or became abruptly larger weakened the impression of human control for two participants. P12 described the movement as ``a little bit too much'' and felt that someone was ``trying to convince me, oh, I'm a human, I'm moving''. P11 similarly described a VH that had ``very little movement'' before it ``suddenly [made] a big smile with a big hand gesture'', which appeared ``unnatural''.

The perceived quality of movement also contributed to participants' agency judgements. Four participants used movements described as smooth, organic, flexible, or detailed as evidence of human control. However, smoothness did not necessarily indicate human control. P19 described the movement as ``too smooth'' and said, ``I felt it was a bit like a computer''. Eight participants associated movement described as jittery, stiff, choreographed, pre-designed, robotic, synthetic, or strange with computer control. Here, ``pre-designed'' referred to how formulaic the gestures appeared; P10 described them as ``routine gestures that people in the interview will have''.

Repetitive or patterned movements were cited as evidence of computer control by five participants. These included gestures that repeated or followed a pattern, as well as gestures that remained the same across answers or different VHs. P1, for example, noted that the hand ``opens the same way, like twice somehow''. Conversely, two participants used flexibility or the absence of an apparent repeating animation as evidence of human control.

Participants also assessed whether movements corresponded with what the VH was saying or expressing. Twelve participants used correspondence between gestures and speech or answer content, perceived emotion, facial expression, or eye behaviour as evidence of human control. P14, for example, assessed whether a VH used ``the right motion to describe things'', while P18 described body language that matched eyebrow movement and head shaking as ``a natural human combination''. However, synchrony alone did not necessarily indicate human control. P8 described the gestures and timing as ``extremely synced'' and therefore ``too good to be true''. Conversely, mismatches between movement and speech, emotion, or another nonverbal behaviour supported computer control for four participants. P11 explained that a gesture appeared robotic because ``that gesture is not being used particularly in that motion'', while P18 felt that ``her face expression doesn't really match her body language''.

The greeting wave was a specific movement discussed by 12 participants. Ten participants used it as evidence of human control, while five used it as evidence of computer control. P6, for example, said that the wave made the VH seem ``a lot more engaged''. P13 instead perceived the greeting as unnatural and described it as ``just like a robot'', whereas P15 noted that the VHs did not acknowledge whether or when P15 returned their waves: ``no one was reacting to me in real time''. Among them, three participants made comparisons across VHs and used differences in waving behaviour as evidence in both directions. Their judgements depended on whether the VH waved first or responded only after a delay, whether the VH waved back, and whether the greeting appeared natural or synthetic.

\section{Scripts}

\includegraphics[width=\textwidth]{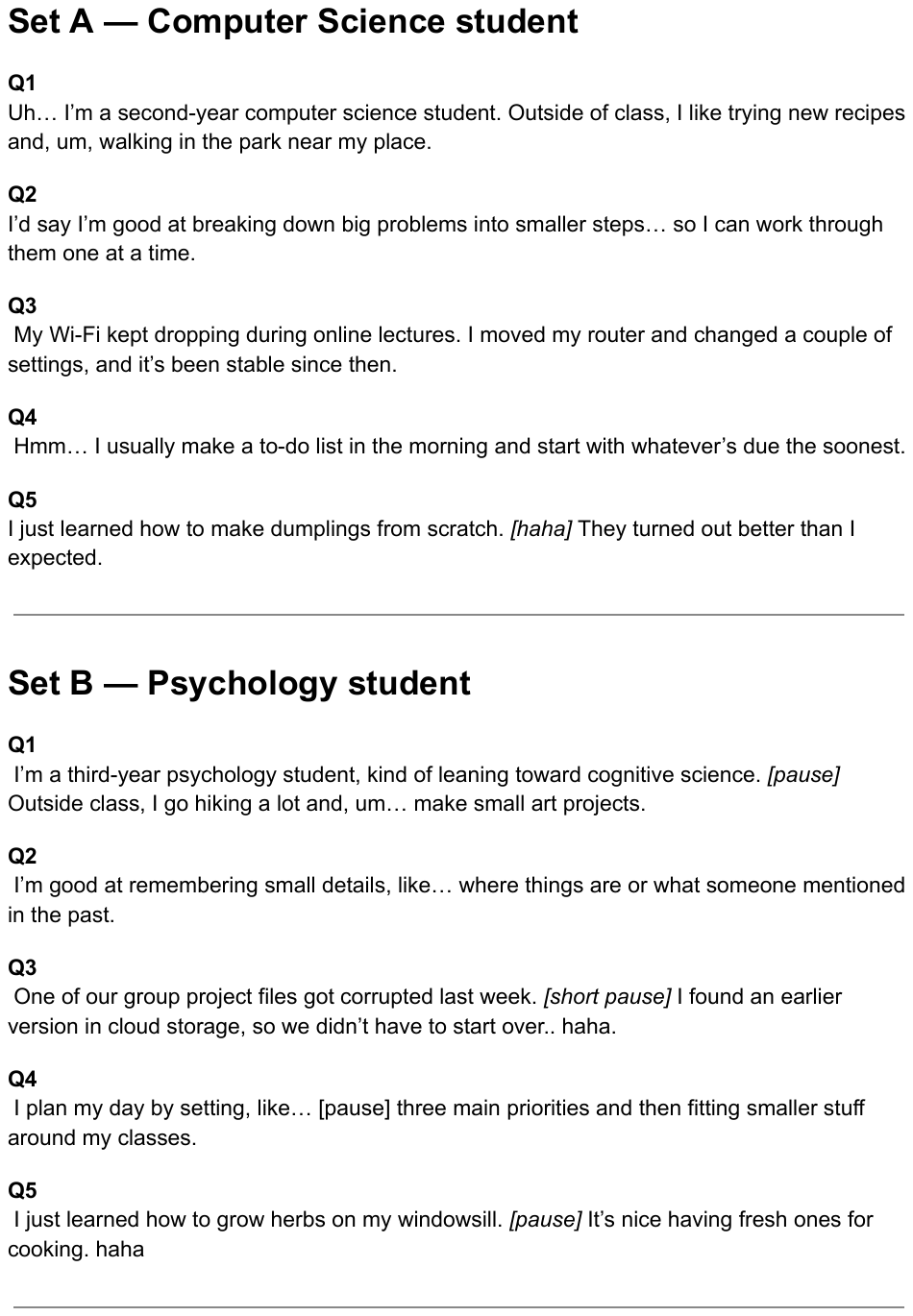}

\newpage
\includegraphics[page=2, width=\textwidth]{Scripts_Questions_and_Answers.pdf}

\newpage
\section{Questionnaire}

\includegraphics[width=\textwidth]{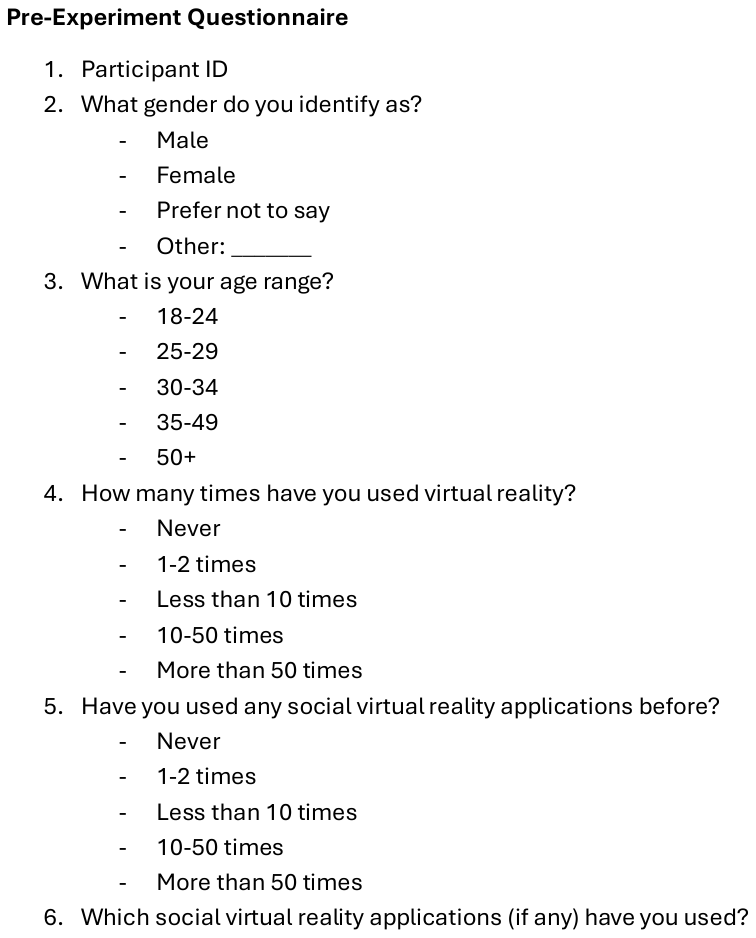}

\newpage
\includegraphics[page=2, width=\textwidth]{Questionnaires.pdf}

\newpage
\includegraphics[page=3, width=\textwidth]{Questionnaires.pdf}

\newpage
\section{Counterbalanced Order}

\includegraphics[angle=90,width=\textwidth]{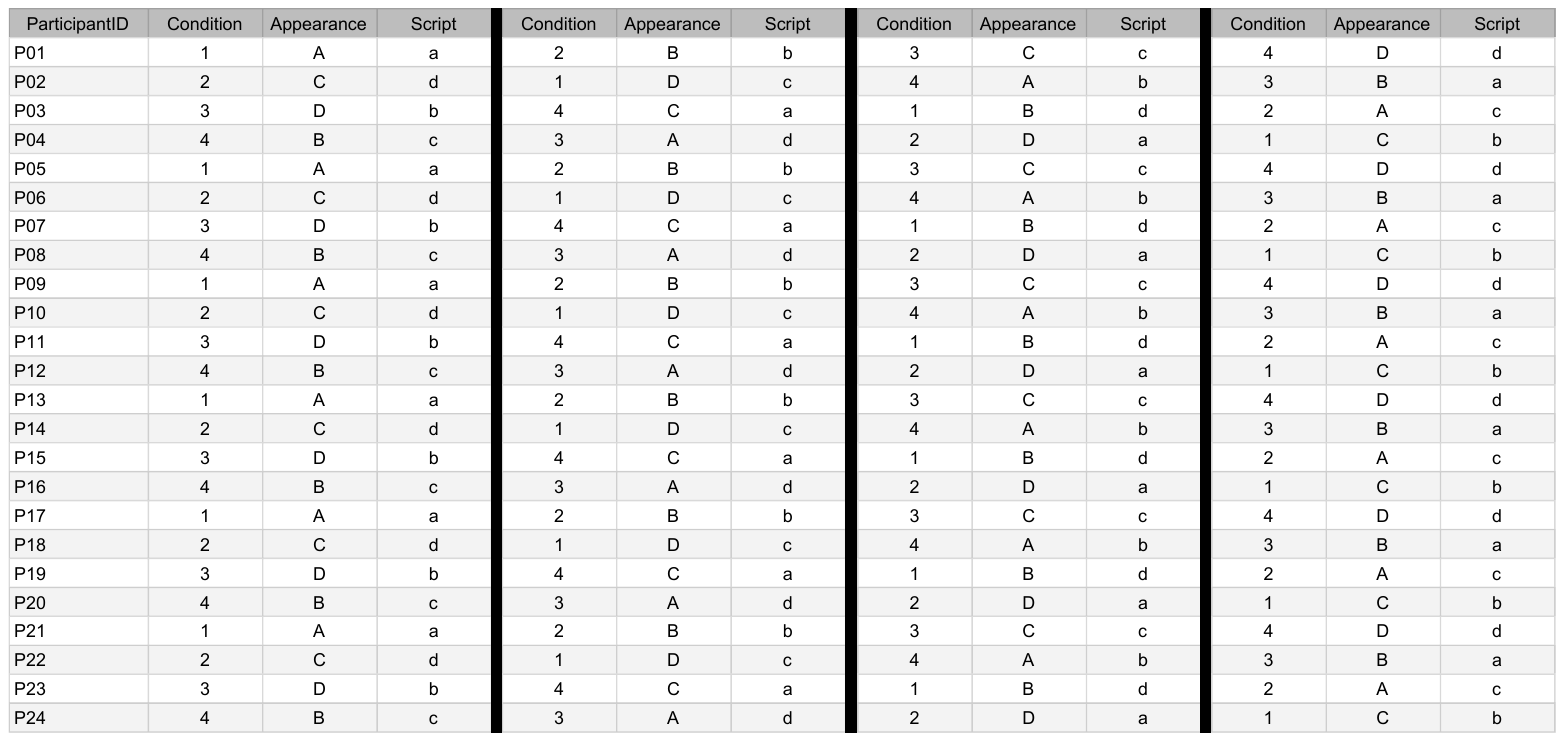}